\documentclass[epj]{svjour}
\usepackage{graphicx}

\newcommand{\Exc}{\Delta\! \nu}

\newcommand{\U}[1]{\ensuremath{\mathrm{\,#1}}}

\begin{document}
\title{Three dimensional cooling and trapping with a narrow line}
\author{T. Chaneli\`ere\inst{1}, L. He\inst{2}, R. Kaiser\inst{3} and D. Wilkowski\inst{3}}
\institute{ $^+$ Now at: Laboratoire Aim\'e Cotton, CNRS, UPR 3321,
Universit\'e Paris-Sud, Bat. 505, F-91405 Orsay Cedex, France \and
$^*$ Now at: State Key Laboratory of Magnetic Resonance and Atomic
and Molecular Wuhan Institute of Physics and Mathematics, Chinese
Academy of Sciences, Wuhan 430071, P. R. China \and Institut Non
Lin\'eaire de Nice, CNRS, UMR 6618, Universit\'e de Nice
Sophia-Antipolis, F-06560 Valbonne, France.}
\date{\today{}}

\abstract{The intercombination line of Strontium at 689nm is
successfully used in laser cooling to reach the photon recoil limit
with Doppler cooling in a magneto-optical traps (MOT). In this paper
we present a systematic study of the loading efficiency of such a
MOT. Comparing the experimental results to a simple model allows us
to discuss the actual limitation of our apparatus. We also study in
detail the final MOT regime emphasizing the role of gravity on the
position, size and temperature along the vertical and horizontal
directions. At large laser detuning, one finds an unusual situation
where cooling and trapping occur in the presence of a high bias
magnetic field.}

 \maketitle

\section{Introduction}

Cooling and trapping alkaline-earth atoms offer interesting
alternatives to alkaline atoms. Indeed, the singlet-triplet
forbidden lines can be used for optical frequency measurement and
related subjects \cite{metro}. Moreover, the spinless ground state
of the most abundant bosonic isotopes can lead to simpler or at
least different cold collisions problems than with alkaline atoms
\cite{coll}. Considering fermionic isotopes, the long-living and
isolated nuclear spin can be controlled by optical means \cite{NMR}
and has been proposed to implement quantum logic gates \cite{infoQ}.
It has also been shown that the ultimate performance of Doppler
cooling can be greatly improved by using narrow transitions whose
photon recoil frequency shifts $\omega_r$ are larger than their
natural widths $\Gamma$ \cite{Castin}. This is the case for the
$^1S_0\rightarrow ^3P_1$ spin-forbidden line of Magnesium
($\omega_r\approx 1100\Gamma$) or Calcium ($\omega_r\approx
36\Gamma$). Unfortunately, both atomic species can not be hold in a
standard magneto-optical trap (MOT) because the radiation pressure
force is not strong enough to overcome gravity. This imposes the use
of an extra quenching laser as demonstrated for Ca \cite{Calcium}.
For Strontium, the natural width of the intercombination transition
($\Gamma=2\pi\times 7.5\U{kHz}$) is slightly broader than the recoil
shift ($\omega_r=2\pi\times 4.7\U{kHz}$). The radiation pressure
force is higher than the gravity but at the same time the final
temperature is still in the $\mu K$ range \cite{Katori99,Jila04}. In
parallel, the narrow transition partially prevents multiple
scattering processes and the related atomic repulsive force
\cite{sesko91}. Hence important improvements on the spatial density
have been reported \cite{Katori99}. However, despite experimental
efforts, such as adding an extra confining optical potential, pure
optical methods have not allowed yet to reach the quantum degeneracy
regime with Strontium atoms \cite{Katori00}.

In this paper, we will discuss some performances, essentially in
terms of temperatures, sizes and loading rates, of a Strontium 88
MOT using the $689\U{nm}$ $^{1}S_{0} \rightarrow ^{3}\!\!P_{1}$
intercombination line.

Initially the atoms are precooled in a MOT on the spin-allowed
$461\U{nm}$ $^{1}S_{0}\rightarrow ^{1}\!\!P_{1}$ transition (natural
width $\Gamma=2\pi\times 32\U{MHz}$) as discussed in
\cite{ChaneliereJosaB}. Then the atoms are transferred into the
$689\U{nm}$ intercombination MOT. To achieve a high loading rate,
Katori et al. \cite{Katori99} have used laser spectrum, broadened by
frequency modulation. Thus the velocity capture range of the
$689\U{nm}$ MOT matches the typical velocity in the $461\U{nm}$ MOT.
They report a transfer efficiency of $30\%$. The same value of
transfer efficiency is also reported in reference \cite{Jila04}. In
our set-up, $50\%$ of the atoms initially in the blue MOT are
transferred into the red one. In section \ref{SecLoading} we present
a systematic study of the transfer efficiency as function of the
parameters of the frequency modulation. In order to discuss the
intrinsic limitations of the loading efficiency, we compare our
experimental results to a simple model. In particular, we
demonstrated that our transfer efficiency is limited by the size of
the red MOT beams. We show that it could be optimized up to $90\%$
with realistic laser power ($25\U{mW}$ per beams).

The minimum temperature achieved in the broadband MOT is about
$2.5\U{\mu K}$. In order to reduce the temperature down to the
photon recoil limit ($0.5\U{\mu K}$), we apply a second cooling
stage, using a single frequency laser and observe similar
temperatures, detuning and intensity dependencies as reported in the
literature (see references \cite{Katori99}, \cite{Jila04},
\cite{LoftusPRA} and \cite{vogel}). In those publications, the role
of gravity on the cooling and trapping dynamics along the $z$
vertical direction has been discussed. In this paper we compare the
steady state behaviour along vertical ($z$) direction to that along
the horizontal plane ($x-y$) where gravity plays indirectly a
crucial role (section \ref{SecCooling}).

Details about the dynamics are given in references
\cite{Jila04},\cite{LoftusPRA}. In particular the authors establish
three regimes. In regime (I) the laser detuning $|\delta|$ is larger
than the power-broadened linewidth $\Gamma_E=\Gamma\sqrt{(1+s)}$.
Regime (II) on the contrary corresponds to $\Gamma_E>|\delta|$. In
both regimes (I) and (II) $\Gamma_E\gg\Gamma,\omega_r$ and the
semiclassical limit is a good approximation. In regime (III) the
saturation parameter is small and a full quantum treatment is
required. We will focus here on the semiclassical regime (I). In
this regime, we confirm that the temperature along the $z$ direction
is independent of the detuning $\delta$. Following Loftus et al.
\cite{LoftusPRA}, we have also found (see section \ref{Vertical
direction}) that this behavior is due to the balance of the
gravitational force and the radiation pressure force produced by the
upward pointing laser (the gravity defining the downward direction).
The center of mass of the atomic cloud is shifted downward from the
magnetic field quadrupole center. As a consequence, cooling and
trapping in the horizontal plane occur at a strong bias magnetic
field mostly perpendicular to the cooling plane. This unusual
situation is studied in detail (section \ref{horizontal plane}).
Despite different friction and diffusion coefficients along the
horizontal and the vertical directions, the horizontal temperature
is found to be the same as the vertical one (see section
\ref{verthory}). In reference \cite{LoftusPRA}, the trapping
potential is predicted to have a box shape whose walls are given by
the laser detuning. This is indeed the case without a bias magnetic
field along the $z$ axis. It is actually different for the regime
(I) described in this paper. Here we have found that the trapping
potential remain harmonic. This leads to a cloud width in the
horizontal direction which is proportional to $\sqrt{|\delta|}$
(section \ref{horizontal plane}).

\section{Experimental set-up}
\label{LaserLocking}

\begin{figure}[h!]
\includegraphics[width=8cm]{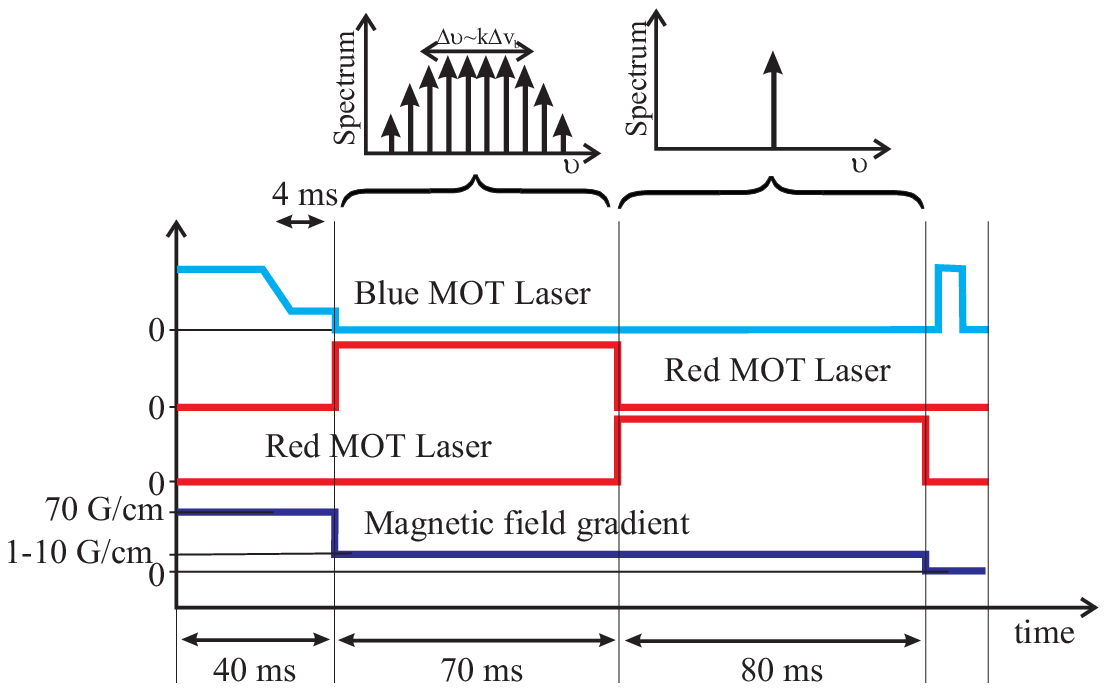}
\caption{Time sequence and cooling stages of Strontium with the
dipole-allowed transition and with the intercombination line.}
\label{LoadRed}
\end{figure}

Our blue MOT setup (on the broad $^{1}S_{0} \rightarrow
^{1}\!\!P_{1}$ transition at $461\U{nm}$) is described in references
\cite{BidelPRL,bruce02}. Briefly, it is composed by six independent
laser beams typically $10 \U{mW/cm^2}$ each. The magnetic field
gradient is about $70\U{G/cm}$. The blue MOT is loaded from an
atomic beam extracted from an oven at $550\U{^{\circ} C}$ and
longitudinally slowed down by a Zeeman slower. The loading rate of
our blue MOT is of $10^9 \U{atoms/s}$ and we trap about $2.10^6$ in
a $0.6\U{mm}$ rms radius cloud when no repumping lasers are used
\cite{Wilkowski05}. To optimize the transfer into the red MOT, the
temperature of the blue MOT should be as small as possible. As
previously observed \cite{ChaneliereJosaB}, this temperature depends
strongly on the optical field intensity. We therefore decrease the
intensity by a factor 5 (see figure \ref{LoadRed}) $4\U{ms}$ before
switching off the blue MOT. The rms velocity right before the
transfer stage is thus reduced down to $\sigma_b = 0.6\U{m/s}$
whereas the rms size remains unchanged. Similar two stage cooling in
a blue MOT is also reported in reference \cite{vogel}.

The $689\U{nm}$ laser source is an anti-reflection coated laser
diode in a $10\U{cm}$ long extended cavity, closed by a diffraction
grating. It is
 locked to an ULE cavity using the Pound-Drever-Hall technique
\cite{LockFlo}. The unity gain of the servo loop is obtained at a
frequency of $1\U{MHz}$. From the noise spectrum of the error
signal, we derive a frequency noise power. It shows, in the range of
interest, namely $1\U{Hz}-100\U{kHz}$, an upper limit of 160
$\U{Hz^2/Hz}$ which is low enough for our purpose. The transmitted
light from the ULE cavity is injected into a $20\U{mW}$ slave laser
diode. Then the noise components at frequencies higher than the ULE
cavity cut-off ($300\U{kHz}$) are filtered. It is important to note
that the lateral bands used for the lock-in are also removed. Those
lateral bands, at $20\U{MHz}$ from the carrier, are generated
modulating directly the current of the master laser diode. A
saturated spectroscopy set-up on the $^{1}S_{0} \rightarrow
^{3}\!\!P_{1}$ intercombination line is used to compensate the long
term drift of $10-50\U{Hz/s}$ mainly due to the daily temperature
change of the ULE cavity.

The slave beam is sent through an acousto-optical modulator mounted
in a double pass configuration. The laser detuning can then be tuned
within the range of a few hundreds of linewidth around the
resonance. This acousto-optical modulator is also used for frequency
modulation (FM) of the laser, as required during the loading phase
(see section \ref{SecLoading}).

The red MOT is made of three retroreflected beams with a waist of
$0.7\U{cm}$. The maximum intensity per beam is about $4\U{mW/cm^2}$
(the saturation intensity being $I_s=3\U{\mu W/cm^2}$). The magnetic
gradient used for the red MOT is varied from $1$ to $10\U{G/cm}$.

To probe the cloud (number of atoms and temperature) we use a
resonant $40\U{\mu s}$ pulse of blue light (see fig \ref{LoadRed}).
The total emitted fluorescence is collected onto an avalanche
detector. From this measurement, we deduce the number of atoms and
then evaluate the transfer rate into the red MOT. At the same time,
an image of the cloud is taken with an intensified CCD camera. The
typical spatial resolution of the camera is $30\U{\mu m}$. Varying
the dark period (time-of-flight) between the red MOT phase and the
probe, we get the ballistic expansion of the cloud. We then derive
the velocity rms value and the corresponding temperature.

\section{Broadband loading of the red MOT}
\label{SecLoading}

The loading efficiency of a MOT depends strongly on the width of the
transition. With a broad transition, the maximum radiation pressure
force is typically $a_m=\frac{v_r\Gamma}{2}\approx 10^4\times g$,
where $v_r$ is the recoil velocity \cite{metcalf}. Hence, on
$l\approx 1\U{cm}$ (usual MOT beam waist) an atom with a velocity
$v_c=\sqrt{2a_ml}\approx 30\U{m/s}$ can be slowed down to zero and
then be captured. During the deceleration, the atom remains always
close to resonance because the Doppler shift is comparable to the
linewidth. Thus MOTs can be directly loaded from a thermal vapor or
a slow atomic beam using single frequency lasers. Moreover typical
magnetic field gradients of few tens of $\U{G/cm}$ usually do not
drastically change the loading because the Zeeman shift over the
trapping region is also comparable to the linewidth.

An efficient loading is more complex to achieved with a narrow
transition. For Strontium, the maximum radiation pressure force of a
single laser is only $a_m\approx 15\times g$. Assuming the force is
maximum during all the capture process, one gets
$v_c=\sqrt{2a_ml}\approx 1.7\U{m/s}$. Hence, precooling in the blue
MOT is almost mandatory. In that case the initial Doppler shift will
be $v_c\lambda^{-1}\approx 2.5\U{MHz}$, $300$ times larger than the
linewidth. In order to keep the laser on resonance during the
capture phase, the red MOT lasers must thus be spectrally broadened.
Because of the low value of the saturation intensity, the spectral
power density can easily be kept large enough to maintain a maximum
force with a reasonable total power (few milliwatts). The magnetic
field gradient of the MOT may also affect the velocity capture
range. To illustrate this point, let us consider an atom initially
in the blue MOT at the center of the trap with a velocity
$v_c=1.7\U{m/s}$. During the deceleration, the Doppler shift
decreases whereas the Zeeman shift increases. However, the magnetic
field gradient does not affect the capture velocity as far as the
total shift (Doppler+Zeeman) is still decreasing. This condition is
fulfilled if the magnetic field gradient is lower than
\cite{phillips82}:
\begin{equation} b_c=\frac{a_m}{\lambda g_e\mu_bv_c}\approx 0.6\U{G/cm}\label{bmax}\end{equation}
where $g_e=1.5$ is the Land\'e factor of the $^{3}P_{1}$ level and
$\mu_b=1.4\U{MHz/G}$ is the Bohr magneton. In practice we use a
magnetic field gradient which is larger than $b_c$. In that case, it
is necessary to increase the width of the laser spectrum so that the
optimum transfer rate is not limited by the Zeeman shift (see
section \ref{theory}). An alternative solution may consist of
ramping the magnetic field gradient during the loading
\cite{Katori99}.

\subsection{Transfer rate: experimental results}

In this section we will present the experimental results regarding
the loading efficiency of the red MOT from the blue MOT. To optimize
the transfer rate, the laser spectrum is broadened using frequency
modulation (FM). Thus the instantaneous laser detuning is
$\Delta(t)=\delta+\Exc.\sin{\nu_mt}$. $\Exc$ and $\nu_m$ are the
frequency deviation and modulation frequency respectively, $\delta$
is the carrier detuning. Here, the modulation index $\Exc/\nu_m$ is
always larger than 1, thus the so-called wideband limit is well
fulfilled. Hence one can assume the FM spectrum to be mainly
enclosed in the interval $[\delta-\Exc;\delta+\Exc]$.

\begin{figure}[h!]
\includegraphics[width=7cm]{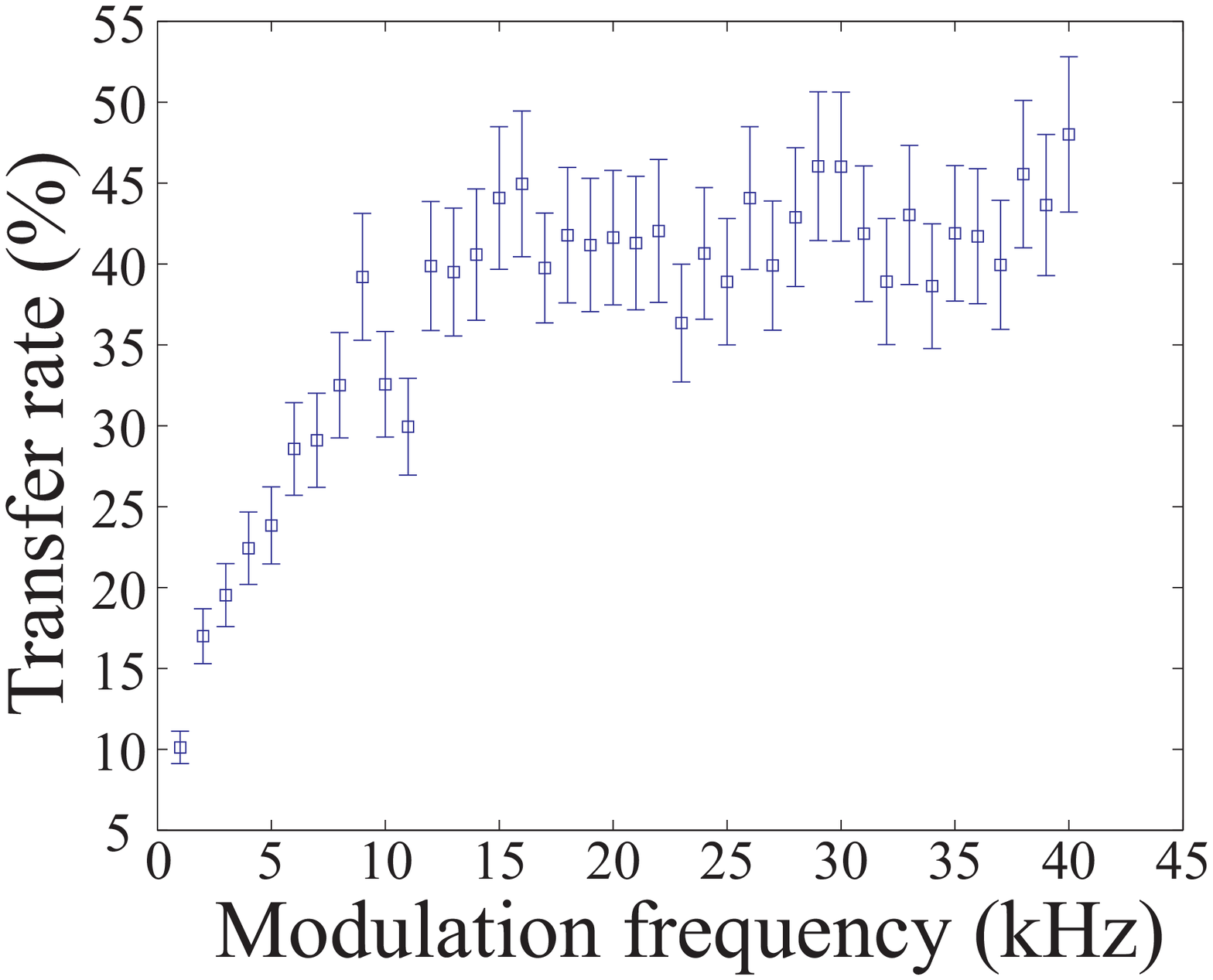}
\caption{Transfer rate as a function of the modulation frequency.
The other parameters are fixed: $P=3\U{mW}$, $\delta=-1000\U{kHz}$,
$b=1\U{G/cm}$ and $\Exc=1000\U{kHz}$} \label{TvsOmegaM}
\end{figure}

As shown in figure \ref{TvsOmegaM}, the transfer rate increases with
$\nu_m$ up to $15\U{kHz}$ where we observe a plateau at $45\%$
transfer efficiency. On the one hand when $\nu_m$ is larger than the
linewidth, the atoms are in the non-adiabatic regime where they
interact with all the Fourier components of the laser spectrum.
Moreover, the typical intensity per Fourier component remains always
higher than the saturation intensity $I_s=3\U{\mu W/cm^2}$. As a
consequence, the radiation pressure force should be close to its
maximum value for any atomic velocity. On the other hand when
$\nu_m<\Gamma/2\pi$, the atoms interact with a chirped intense laser
where the mean radiation pressure force (over a period $2\pi/\nu_m$)
is clearly smaller than in the case $\nu_m>\Gamma/2\pi$. As a
consequence, the transfer rate is reduced when $\nu_m$ decreases.

In figure \ref{TvsExc}, the transfer rate is measured as a function
of $\Exc$. The carrier detuning is $\delta=-1\U{MHz}$ and the
modulation frequency is kept larger than the linewidth
($\nu_m=25\U{kHz}$). Starting from no deviation ($\Exc=0$), we
observe (fig. \ref{TvsExc}) an increase of the transfer rate with
$\Exc$ (in the range $0<\Exc<500\U{kHz}$). After reaching its
maximum value, the transfer rate does not depend on $\Exc$ anymore.
Thus the capturing process is not limited by the laser spectrum
anymore. If we further increase the frequency deviation $\Exc$, the
transfer becomes less efficient and finally decreases again down to
zero. This reduction occurs as soon as $\Exc>|\delta|$, \emph{i.e.}
some components of the spectrum are blue detuned. This frequency
configuration obviously should affect the MOT steady regime adding
extra heating at zero velocity (see section \ref{broa-Temp}). We can
see that it is also affecting the transfer rate. To confirm that
point, figure \ref{TvsExc2} shows the same experiment but with a
larger detuning $\delta=-1.5\U{MHz}$ and $\delta=-2\U{MHz}$ for the
figures \ref{TvsExc2}a and \ref{TvsExc2}b respectively. Again the
transfer rate decreases as soon as $\Exc>|\delta|$. The transfer
rate is also very small on the other side for small values of
$\Exc$. In that case the entire spectrum of the laser is too far red
detuned. The radiation pressure forces are significant only for
velocities larger than the capture velocity and no steady state is
expected. Keeping now the deviation fixed and varying the detuning
as shown in figure \ref{Tvsdelta}, we observe a maximum transfer
rate when the detuning is close to the deviation frequency $\Exc
\simeq |\delta|$. Closer to resonance ($\Exc < |\delta|$), the blue
detuned components prevent an efficient loading of the MOT.

The magnetic field gradient plays also a crucial role for the
loading. We indeed observe (fig. \ref{Tvsb}) that the transfer rate
decreases when the magnetic field gradient increases. At very low
magnetic field ($b<1\U{G/cm}$) the reduction of the transfer rate is
most likely due to a lack of stability within the trapping region.
In that case we actually observe a strong displacement of the center
of mass of the cloud. This is induced by imperfections of the set-up
such as non-balanced laser intensities which are critical at low
magnetic gradient. Hence, the optimum magnetic field gradient is
found to be the smallest one which ensure the stability of the cloud
in the MOT.

\subsection{Theoretical model and comparison with the experiments}
\label{theory}

To clearly understand the limiting processes of the transfer rate,
we compare the experimental data to a simple 1D theoretical model based on the following assumptions:\\
- An atom undergoes a radiation pressure force and thus a
deceleration if the modulus of its velocity is between $v_{max}$ and
$v_{min}$ with
\begin{equation}
v_{max}=\lambda(|\delta|+\Exc),\quad
v_{min}=\max\{\lambda(|\delta|-\Exc);\lambda(-|\delta|+\Exc)\}\label{vminmax}
\end{equation}
$a_m=0$ elsewhere. We simply write that the Doppler shift is
contained within the FM spectrum. We add the condition
$v_{min}=\lambda(-|\delta|+\Exc)$ when some components are blue
detuned $\Exc>|\delta|$. In this case, we consider the simple
ideal situation where the two counter-propagating lasers are assumed perfectly balanced and then compensate each other in the spectral overlapping region.\\
- Even in the semiclassical model, it is difficult to calculate the
acceleration as a function of the velocity for a FM spectrum.
However for all the data presented here, the saturation parameter is
larger than one. Hence the deceleration is set to a constant value
$-\frac{1}{3}a_m$ when $v_{min}<| v|<v_{max}$. The prefactor $1/3$
takes into account the saturation by the 3
counter-propagating laser beam pairs.\\
- The magnetic field gradient is included by giving a spatial
dependence of the
detuning $\delta$ in the expression (\ref{vminmax}).\\
- An atom will be trapped if its velocity changes of sign
within a distance shorter that the beam waist.\\

\begin{figure}[h!]
\includegraphics[width=7cm]{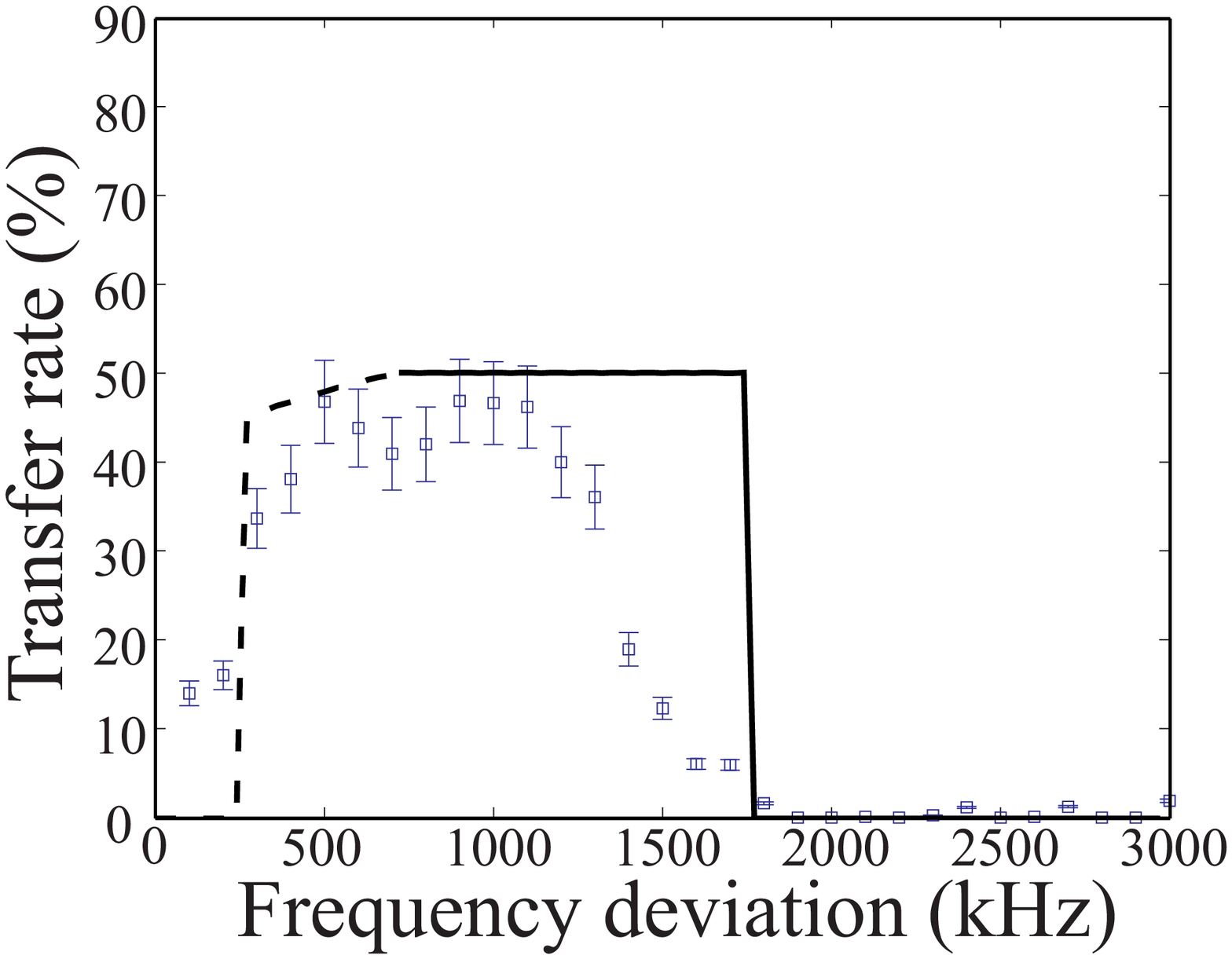}
\caption{Transfer rate as a function of the frequency deviation
(squares). The other parameters are fixed: $P=3\U{mW}$,
$\delta=-1000\U{kHz}$, $b=1\U{G/cm}$ and $\nu_m=25\U{kHz}$. The dash
and solid line correspond to a simple model prediction (see text).
The transfer rate is limited by the frequency deviation of the broad
laser spectrum for the dash line and by the waist of the MOT beam
for the solid line. } \label{TvsExc}
\end{figure}

\begin{figure}[h!]
\includegraphics[width=7cm]{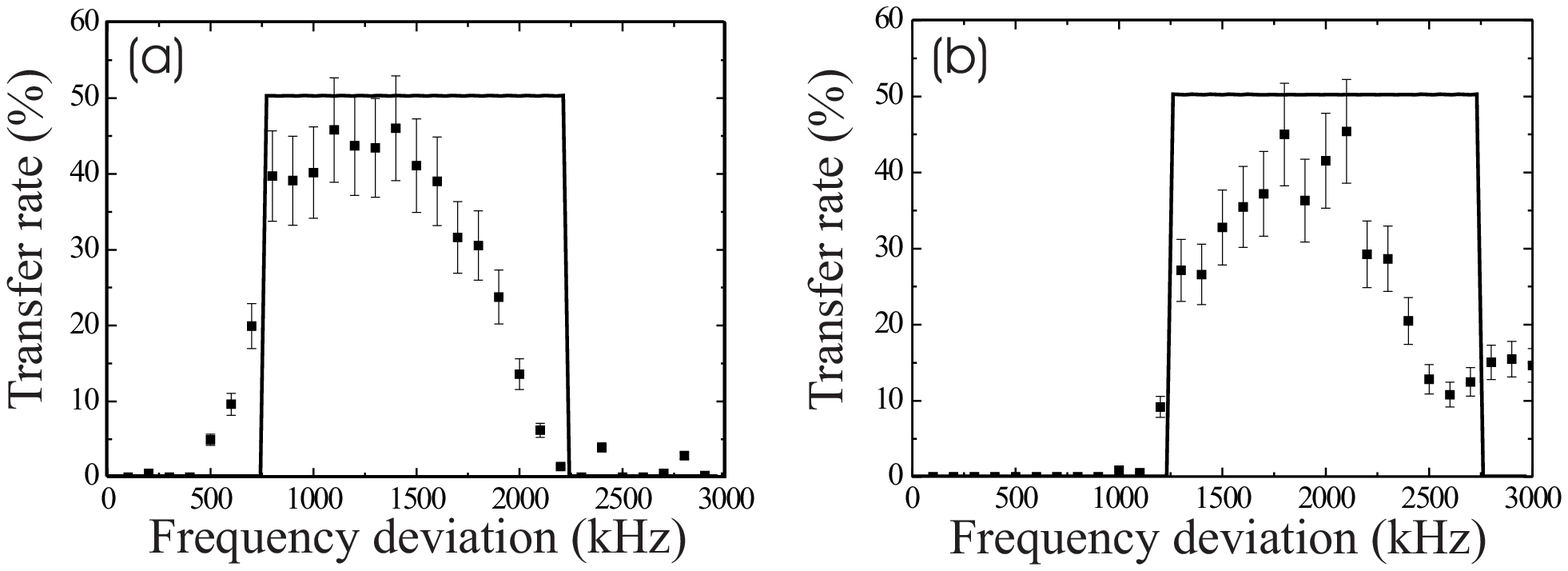}
\caption{Transfer rate as a function of the frequency deviation
(squares). $\delta=-1500\U{kHz}$ and $\delta=-2000\U{kHz}$ for (a)
and (b) respectively, the other parameters and the definitions are
the same than for figure \ref{TvsExc}.} \label{TvsExc2}
\end{figure}

\begin{figure}[h!]
\includegraphics[width=7cm]{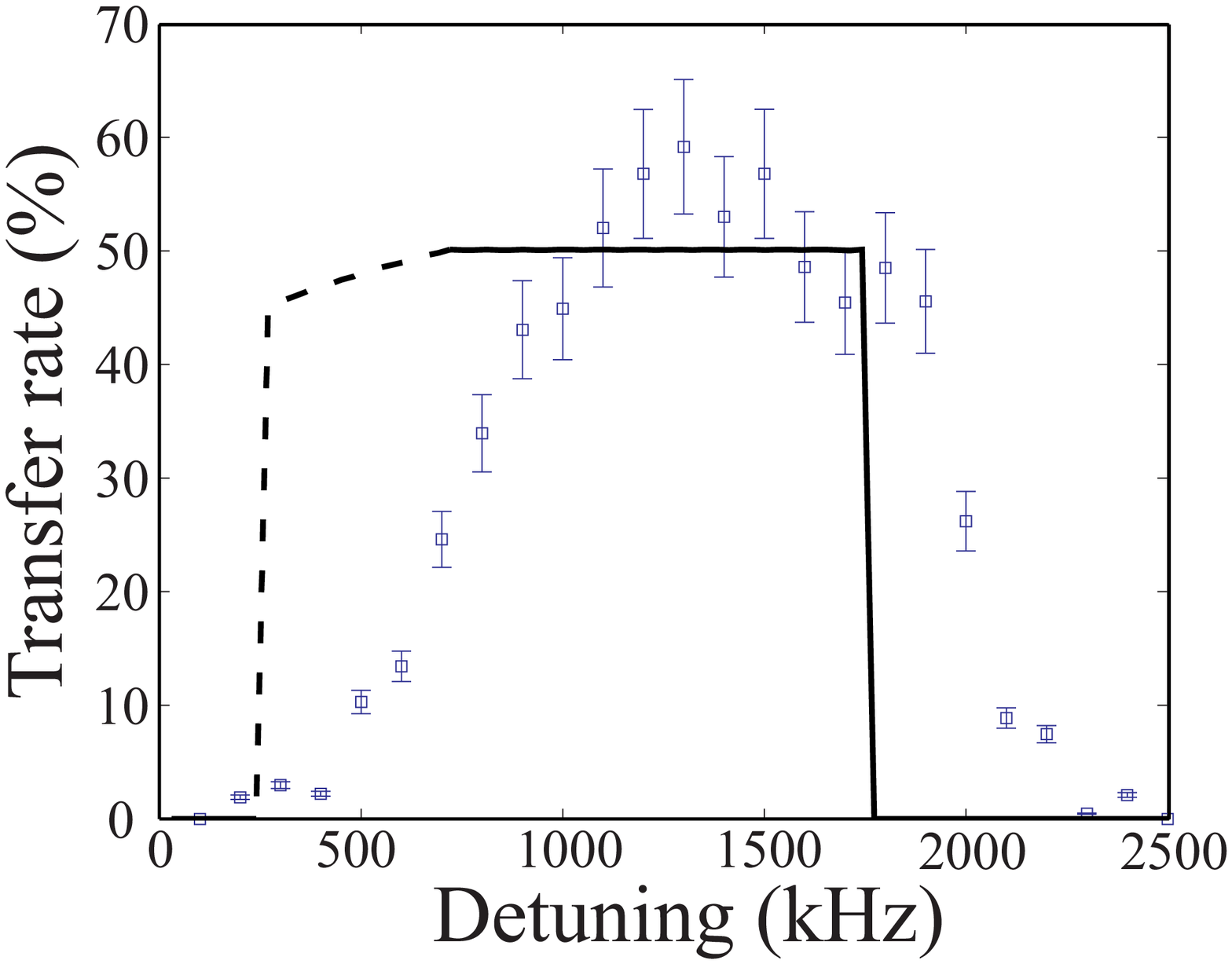}
\caption{Transfer rate as a function of the detuning (squares). The
other parameters are fixed: $P=3\U{mW}$, $\Exc=1000\U{kHz}$,
$b=1\U{G/cm}$ and $\nu_m=25\U{kHz}$. The dashed and solid lines have
the same signification than in figure \ref{TvsExc}.}
\label{Tvsdelta}
\end{figure}

\begin{figure}[h!]
\includegraphics[width=7cm]{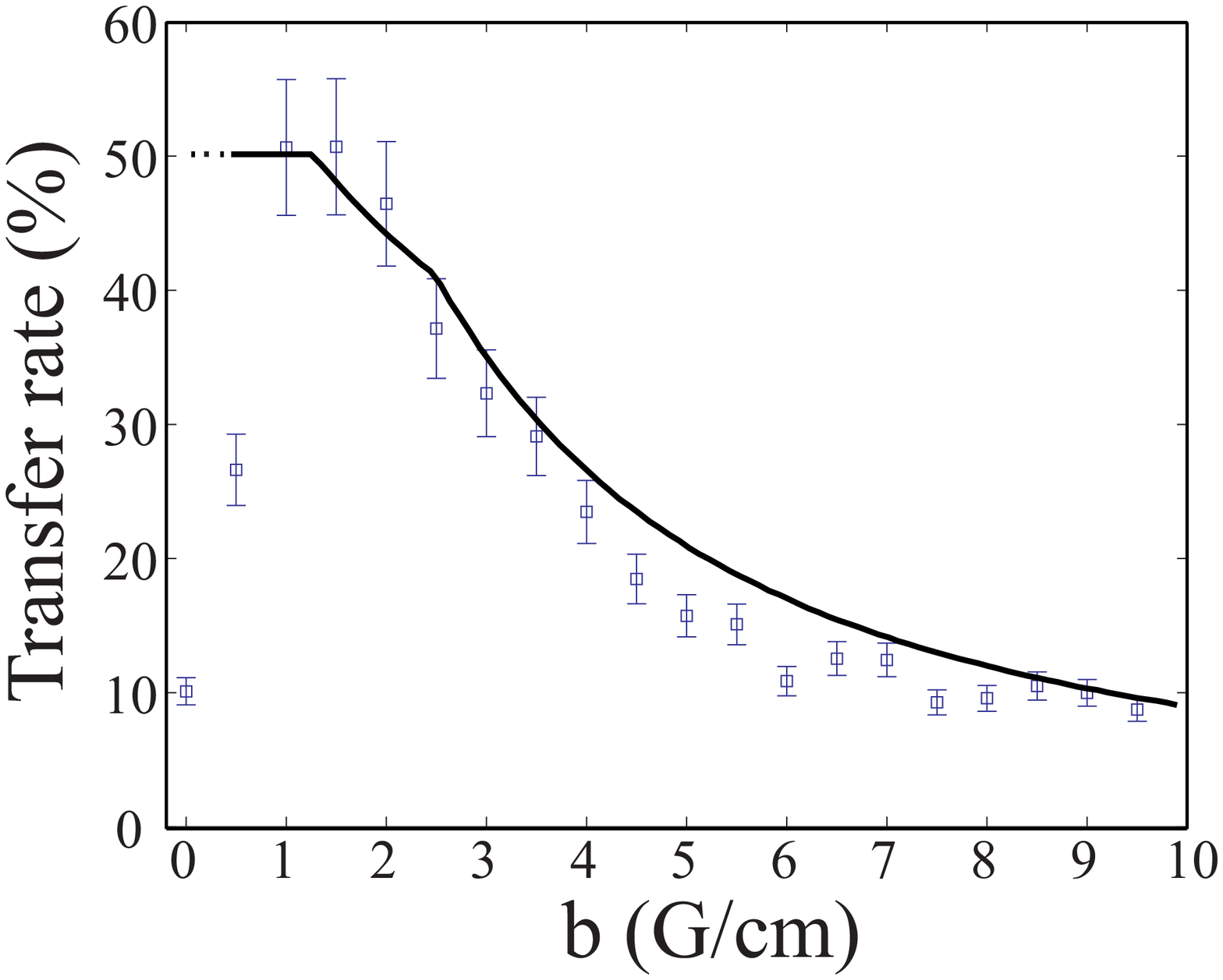}
\caption{Transfer rate as a function of the magnetic gradient
(squares). The other parameters are fixed: $P=3\U{mW}$,
$\delta=-1000\U{kHz}$, $\Exc=1000\U{kHz}$ and $\nu_m=25\U{kHz}$. The
transfer rate is limited by the waist of the MOT beam for all
values. The dotted lines represent the case where the magnetic field
gradient do not affect the deceleration.} \label{Tvsb}
\end{figure}

In figures \ref{TvsExc}-\ref{Tvsb} the results of the model are
compared to the experimental data. The agreement between the model
and the experimental data is correct except at large frequency
deviation (figures \ref{TvsExc} and \ref{TvsExc2}) or at low
detuning (figure \ref{Tvsdelta}). In those cases the spectrum has
some blue detuned components. As mentionned before, this is a
complex situation where the assumptions of the simple model do not
hold anymore. Fortunately those cases do not have any practical
interest because they do not correspond to the optimum transfer
efficiency.

At the optimum, the model suggests that the transfer is limited by
the beam waist (see caption of figures (\ref{TvsExc}-\ref{Tvsb})).
Moreover for all the situation explored in figures
\ref{TvsExc}-\ref{Tvsdelta}, the magnetic field gradient is strong
enough ($b=1\U{G/cm}$) to have an impact on the capture process, as
suggested by the inequality (\ref{bmax}). However it is not the
transfer limiting factor because the Zeeman shift is easily
compensated by a larger frequency excursion or by a larger detuning.

\begin{figure}[h!]
\includegraphics[width=7cm]{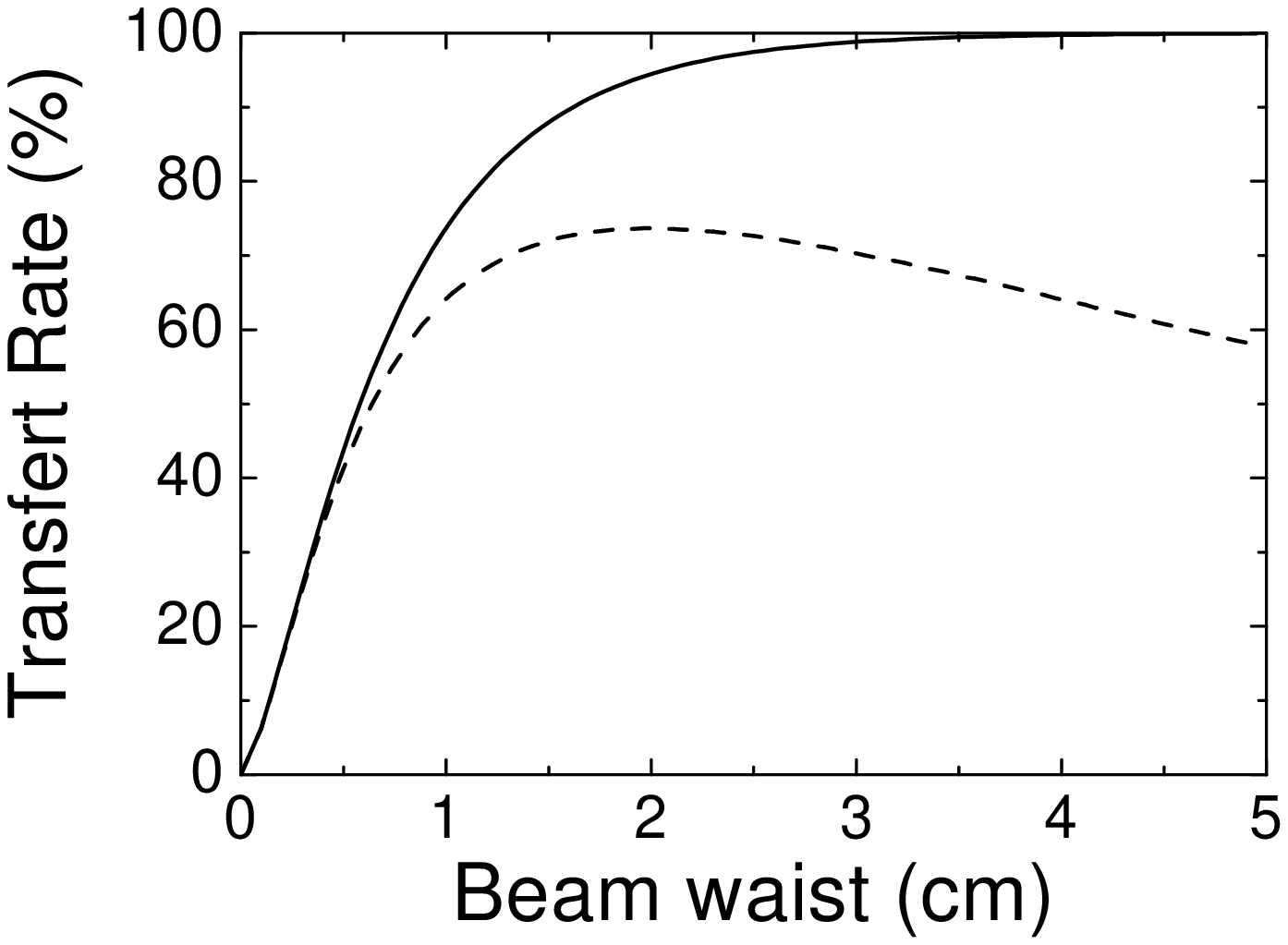}
\caption{Transfer rate as a function of the beam waist. The solid
lines correspond to a high saturation parameter where as the dash
line correspond to a constant power of $P=3\U{mW}$. The other
parameters are fixed: $\delta=-1000\U{kHz}$, $\Exc=1000\U{kHz}$ and
$b=0.1\U{G/cm}$.} \label{Trans_theo}
\end{figure}

Increasing the beam waist would definitely improve the transfer
efficiency as showed in figure \ref{Trans_theo}. If the saturation
parameter would remain large for all values of beam waist, more than
$90\%$ of the atoms would be transferred for a $2\U{cm}$ beam waist.
$25\U{mW}$ of power per beam should be sufficient to achieve this
goal. In our experimental set-up, the power is limited to $3\U{mW}$
per beam. So the saturation parameter is necessarily reduced once
the waist is increased. To take this into account and get a more
realistic estimation of the efficiency for larger beams, we replace
the previous acceleration by the expression $a_ms/(1+3s)$, with
$s=I/I_s$ the saturation parameter per beam. In this case, the
transfer efficiency becomes maximum at $70\%$ for a beam waist of
$1.5\U{cm}$.

\subsection{Temperature}
\label{broa-Temp}

\begin{figure}[h!]
\includegraphics[width=7cm]{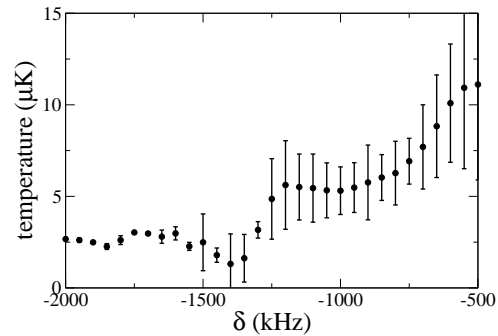}
\caption{Measured temperature as a function of the detuning for a FM
spectrum. The other parameters are fixed: $P=3\U{mW}$,
$b=1\U{G/cm}$, $\Exc=1000\U{kHz}$ and $\nu_m=25\U{kHz}$}.
\label{TempBBSpec}
\end{figure}

Cooling with a broadband FM spectrum on the intercombiaison line
decreases the temperature by three orders of magnitude in comparison
with the blue MOT: from $3 \U{mK}$ ($\sigma_b = 0.6\U{m/s}$) to $2.5
\U{\mu K}$ (see figure \ref{TempBBSpec}). For small detuning, the
temperature is strongly increasing when the spectrum has some blue
detuned components ($\Exc>|\delta|$). Indeed the cooling force and
heating rate are strongly modified at the vicinity of zero detuning.
This effect is illustrated in figure \ref{TempBBSpec}. On the other
side at large detuning ($\delta < - 1.5\U{MHz}$), the temperature
becomes constant. This regime corresponds to a detuning independent
steady state, as also observed in single frequency cooling (see ref.
\cite{LoftusPRA} and section \ref{SecCooling}).

\section{Single frequency cooling}
\label{SecCooling}

About half of the atoms initially in the $461\U{nm}$ MOT are
recaptured in the red one using a broadband laser. The final
temperature is $2.5 \U{\mu K}$ \emph{i.e.} $5$ times larger than the
photon recoil temperature $T_r=460 \U{nK}$. To further decrease the
temperature one has to switch to single frequency cooling (for time
sequences: see figure \ref{LoadRed}). As we will see in this
section, the minimum temperature is now about $600 \U{nK}$ close to
the expected $0.8T_r$ in an 1D molasses \cite{Castin}. Moreover, one
has to note that, under proper conditions described in reference
\cite{LoftusPRA}, the transfer between the broadband and the single
frequency red MOT can be almost lossless.

In the steady state regime of the single frequency red MOT, one has
$k\sigma_v\approx \omega_r\approx\Gamma$. Thus, there is no net
separation of different time scales as in MOTs operated with a broad
transition where $\omega_r<<k\sigma_v<<\Gamma$. However, here the
saturation parameter $s$ always remains high. It corresponds to the
so-called regimes (I) and (II) presented in reference
\cite{LoftusPRA}. Thus $\omega_r<<\Gamma\sqrt{1+s}$ and the
semiclassical Doppler theory describes properly the encountered
experimental situations.

To insure an efficient trapping, the parameter's values of the
single frequency red MOT are different from a usual broad transition
MOT: the magnetic field gradient is higher, typically
$1000\Gamma/cm$. Moreover the gravity is not negligible anymore by
comparison with the typical radiation pressure. Those features lead
to an unusual behavior of the red MOT as we will explain in this
section. We will first independently analyze the MOT properties
along the vertical dimension (section \ref{Vertical direction}) then
in the horizontal plane (section \ref{horizontal plane}), to finally
compare those two situations (section \ref{verthory}).

\subsection{Vertical direction}
\label{Vertical direction}

In the regime (I) \emph{i.e.} at large negative detuning and high
saturation (see examples on figure \ref{TempNLC}a) the temperature
is indeed constant. As explained in reference \cite{LoftusPRA}, this
behavior is due to the balance between the gravity and the radiation
pressure force of the upward laser. At large negative detuning, the
downward laser is too far detuned to give a significant
contribution. In the semiclassical regime, an atom undergoes a net
force of
\begin{equation}
F_z=\hbar
k\frac{\Gamma}{2}\frac{s}{1+s_T+4(\delta-g_e\mu_Bbz-kv_z)^2/\Gamma^2}-mg
\end{equation}
Considering the velocity dependence of the force, the first order
term is:

\begin{equation}
F_z\approx -\gamma_zv_z \label{Fvert}
\end{equation}
with \begin{equation}\gamma_z=-4\frac{\hbar
k^2\delta_{\textrm{\small{eff}}}}{\Gamma}\frac{s}{(1+s_T+4\delta_{\textrm{\small{eff}}}^2/\Gamma^2)^2}\end{equation}
where the effective detuning
$\delta_{\textrm{\small{eff}}}=\delta-g_e\mu_Bb<z>$ is define such
as
\begin{equation}
\hbar
k\frac{\Gamma}{2}\frac{s}{1+s_T+4\delta_{\textrm{\small{eff}}}^2/\Gamma^2}=mg\label{defdeff}\end{equation}
$s_T$ is the total saturation parameter including all the beams.
$<z>$ is the mean vertical position of the cold cloud. Hence
$\delta_{\textrm{\small{eff}}}$ is independent of the laser detuning
$\delta$ and the vertical temperature at larger detuning depends
only on the intensity as shown in figures \ref{TempNLC}a and
\ref{TempNLC}b.

\begin{figure}[h!]
\includegraphics[width=7cm]{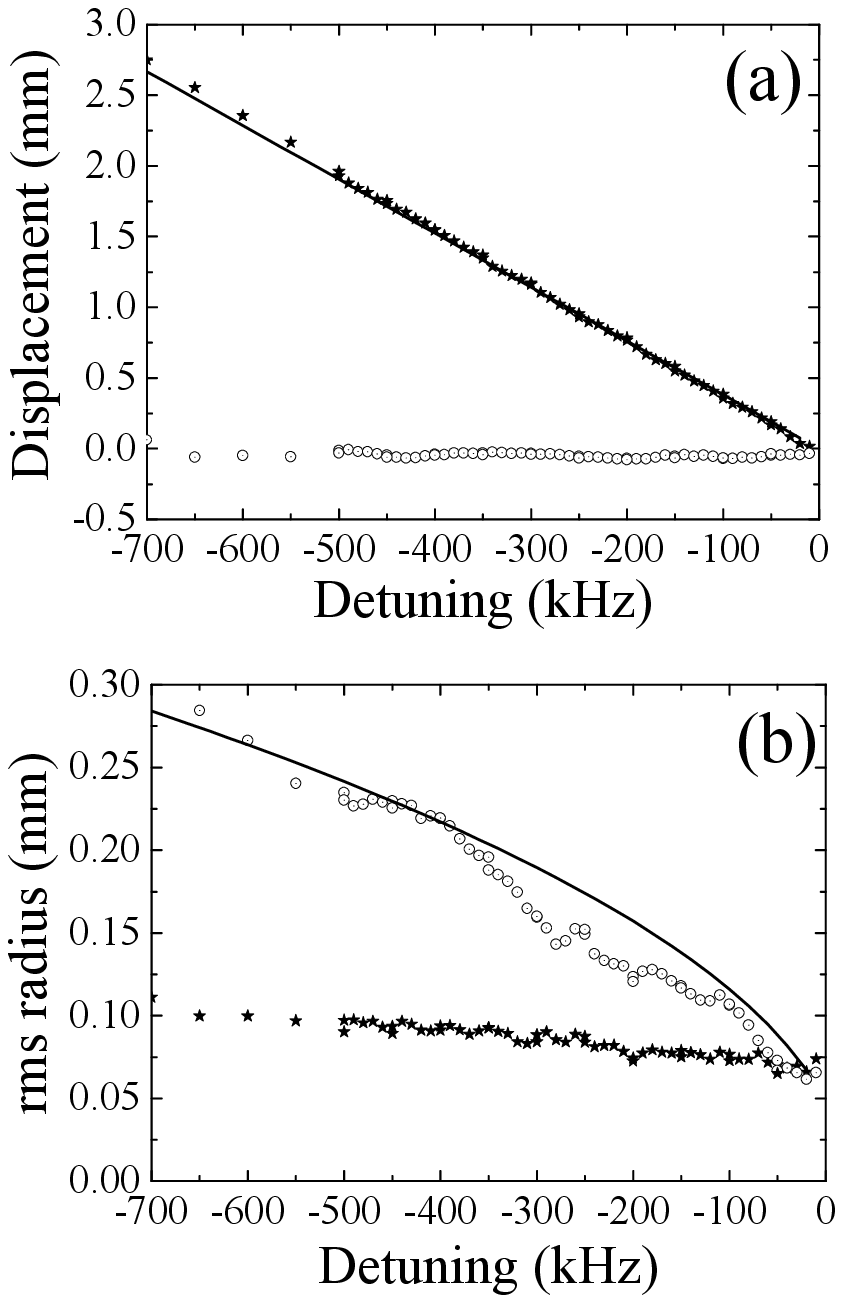}
\caption{Displacement (a) and rms radius (b) of the cold cloud in
single frequency cooling along the $z$ axis (star) and in the $x-y$
plane (circle). The intensity per beam is $I=20 I_s$  and the
magnetic gradient $b=2.5\U{G/cm}$ along the strong axis in the $x-y$
plane. The linear displacement prediction correspond to the plain
line (graph a). In graph b, the plain curve correspond to the rms
radius prediction based on the equipartition theorem.}
\label{RadiusNLC}
\end{figure}

The spatial properties of the cloud are also related to the
effective detuning $\delta_{\textrm{\small{eff}}}$ which is
independent of $\delta$. The mean vertical position depends linearly
on the detuning, so that one has :
\begin{equation}
\frac{\textrm{d}<z>}{\textrm{d}|\delta|}=\frac{-1}{g_e\mu_Bb}
\label{posmoy}
\end{equation}
The predicted vertical displacement is compared to the experimental
data in figure \ref{RadiusNLC}a. The agreement is excellent (the
only adjustable parameter is the unknown origin of the vertical
axe). Because the radiation pressure force for an atom at rest does
not depend on the laser detuning $\delta$, the vertical rms size
should be also $\delta$-independent. This point is also verified
experimentally (see figure \ref{RadiusNLC}b).

\subsection{$x-y$ horizontal plane}
\label{horizontal plane}

\begin{figure}[h!]
\includegraphics[width=5cm]{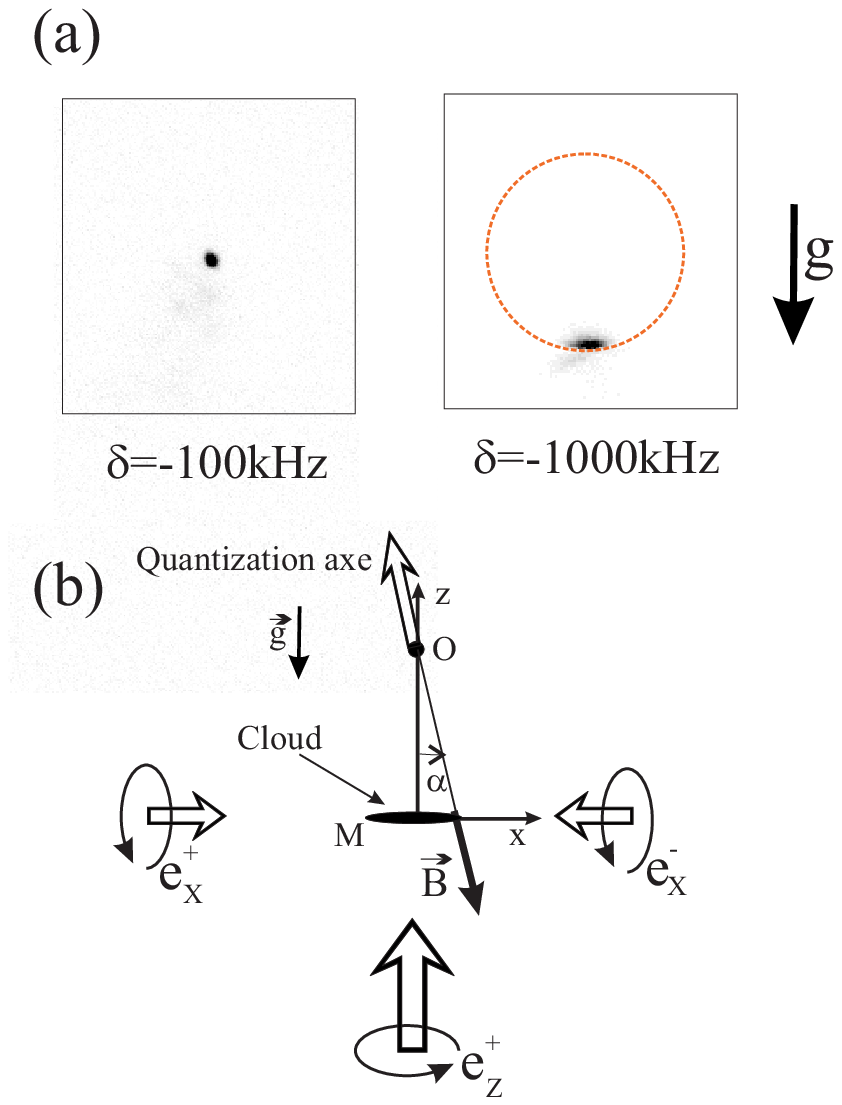}
\caption{(a) Images of the cold cloud in the red MOT. The cloud
position for $\delta=-100\U{kHz}$ coincides roughly with the center
of the MOT whereas it is shifted downward for $\delta=-1000\U{kHz}$.
The spatial position of the resonance correspond dot circle. (b)
Sketch representing the large detuning case. The coupling efficiency
of the MOT lasers is encoded in the size of the empty arrow. The
laser form below has maximum efficiency whereas the one pointing
downward is absent because is too detuned. Along a horizontal axe,
the lasers
 are less coupled because they do not have the
correct polarization. The $\alpha$ angle is the angular position of
an atom $M$ with respect to $O$, the center of the MOT.}
\label{Sch_red_mot}
\end{figure}

Let us now study the behavior of the cold cloud in the $x-y$ plane
at large laser detuning. As explained in section \ref{Vertical
direction}, the position of the cloud is vertically shifted downward
with respect to the center of the magnetic field quadrupole (see
figure \ref{Sch_red_mot}). The dynamic in the $x-y$ plane occurs
thus in the presence of a high bias magnetic field. To derive the
expression of the semiclassical force in this unusual situation one
has first to project the circular polarizations states of the
horizontal lasers on the eigenstates. We define the quantification
axis along the magnetic field, one gets:
\begin{equation}
e^+_{x}=\frac{1+\sin{\alpha}}{2}e^-_{B}+\frac{\cos{\alpha}}{\sqrt{2}}\pi_{B}+\frac{1-\sin{\alpha}}{2}e^+_{B}
\end{equation}
\begin{equation}
e^-_{x}=\frac{1-\sin{\alpha}}{2}e^-_{B}+\frac{\cos{\alpha}}{\sqrt{2}}\pi_{B}+\frac{1+\sin{\alpha}}{2}e^+_{B}
\end{equation}
where $e^-_{i}$, $\pi_i$ and $e^+_i$ represent respectively the
left-handed, linear and right-handed polarisations along the $i$
axis. The angle $\alpha$ between the vertical axis and the local
magnetic field is shown on figure \ref{Sch_red_mot}. For large
detuning, $\alpha$ is always small ($\alpha\ll\,1$ ) and we write
$\alpha\approx -x/<z>$ considering only the dynamics along the $x$
dimension. For simplicity the magnetic field gradient $b$ is
considered as spatially isotropic with $b>0$ as sketched on figure
\ref{Sch_red_mot}b. The expression of the radiation pressure force
is then:
\begin{eqnarray}
F_x=\hbar k\frac{\Gamma}{2}\times \\
(\frac{s(1-\sin{\alpha})^2/4}{1+s_T+4(\delta-g_e\mu_Bb<z>(1-\tan{\alpha})-kv_x)^2/\Gamma^2}
-\nonumber\\
\frac{s(1+\sin{\alpha})^2/4}{1+s_T+4(\delta-g_e\mu_Bb<z>(1-\tan{\alpha})+kv_x)^2/\Gamma^2})\nonumber
\label{PreFac}
\end{eqnarray}
Note that this expression is not restricted to the small $\alpha$
values. We expect six terms in the expression (\ref{PreFac}): three
terms for each laser corresponding to the three $e^-_{B}$, $\pi_B$
and $e^+_B$ polarisation eigenstates. However only two terms,
corresponding to the $e^+_{B}$ state, are close to resonance and
thus have a dominant contribution. As for the vertical dimension,
the off resonant terms are removed from the expression
(\ref{PreFac}). One has also to note that the effective detuning
$\delta_{\textrm{\small{eff}}}=\delta-g_e\mu_Bb<z>$ is actually the
same as the one along the vertical dimension.

The first order expansion of (\ref{PreFac}) in $\alpha$ and
$kv_x/\Gamma$ gives the expression of the horizontal radiation
pressure force:
\begin{equation}
F_x\approx
-\kappa_{\alpha}\alpha-\gamma_xv_x=-\kappa_{x}x-\gamma_xv_x\label{Fx}
\end{equation}
with \begin{equation}\kappa_{\alpha}=-<z>\kappa_x=\hbar
k\frac{\Gamma}{2}
\frac{s}{1+s_T+4\delta_{\textrm{\small{eff}}}^2/\Gamma^2}=mg\end{equation}
and
\begin{equation}\gamma_x=\frac{\gamma_z}{2}=-2\frac{\hbar
k^2\delta_{\textrm{\small{eff}}}}{\Gamma}\frac{s}{(1+s_T+4\delta_{\textrm{\small{eff}}}^2/\Gamma^2)^2}\label{gx}\end{equation}

As for the vertical dimension (equation (\ref{defdeff})), the force
depends on $\delta_{\textrm{\small{eff}}}$ but at the position of
the MOT does not depend on the laser detuning $\delta$. Hence, at
large detuning, the horizontal temperature depends only on the
intensity as observed in figures \ref{TempNLC}a and \ref{TempNLC}b.


To understand the trapping mechanisms in the $x-y$ plane, we now
consider an atom at rest located at a position $x\neq 0$
(corresponding to $\alpha\neq 0$), i.e. not in the center of the
MOT. The transition rate of two counter-propagating laser beam is
not balanced anymore. This is due to the opposite sign in the
$\alpha$ dependency of the prefactor in expression (\ref{PreFac}).
This mechanism leads to a restoring force in the $x-y$ plane at the
origin of the spatial confinement (equation \ref{Fx}). Applying the
equipartition theorem one gets the horizontal $rms$ size of the
cloud:
\begin{equation}
x_{\textrm{\small{rms}}}^2=\frac{k_BT}{\kappa_{x}}=-\frac{<z>k_BT}{mg}
\end{equation}
Without any free adjusting parameter, the agreement with
experimental data is very good as shown in figure \ref{RadiusNLC}b.
On the other hand there's no displacement of the center of mass in
the $x-y$ plane whatever is the detuning $\delta$ as long as the
equilibrium of the counter-propagating beams intensities is
preserved (figure \ref{RadiusNLC}a).

\subsection{Comparing the temperatures along horizontal and vertical axes}
\label{verthory}

\begin{figure}[h!]
\includegraphics[width=7cm]{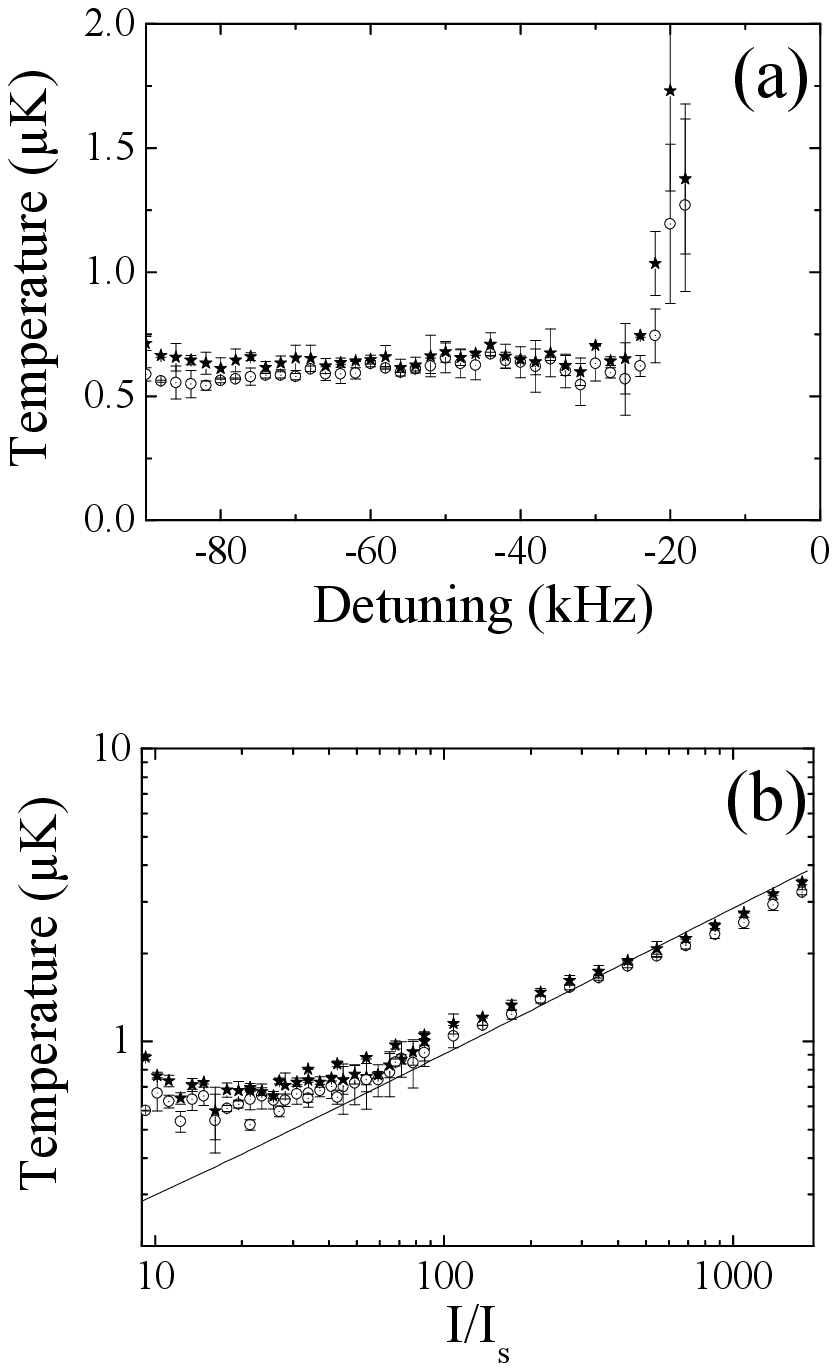}
\caption{Measured temperature as a function of the detuning (a) with
$I=4I_s$ or $I=15I_s$ and as a function of the intensity (b) with
$\delta=-100\U{kHz}$ of single frequency cooling. The circles
(respectively stars) correspond to temperature along one of the
horizontal (respectively vertical) axis. The magnetic field gradient
is $b=2.5\U{G/cm}$}. \label{TempNLC}
\end{figure}

As seen in sections \ref{Vertical direction} and \ref{horizontal
plane}, gravity has a dominant impact on cooling in a MOT operated
on the intercombination line not only along the vertical axe but
also in the horizontal plane. Even so we expect different behaviors
along this directions essentially because the gravity renders the
trapping potential anisotropic. This is indeed the case for the
spatial distribution (figures \ref{RadiusNLC}a and \ref{RadiusNLC}b)
whereas the temperatures are surprisingly the same (figures
\ref{TempNLC}a and \ref{TempNLC}b). We will now give few simple
arguments to physically explain this last point.

In the semiclassical approximation, the temperature is defined as
the ratio between the friction and the diffusion term:
\begin{equation}
k_BT_i=\frac{\gamma_i}{D_i^{abs}+D_i^{spo}}\quad\textrm{with}\quad
i=x,y,z
\end{equation}
$D^{abs}$ and $D^{spo}$ correspond to the diffusion coefficients
induced by absorption and spontaneous emission events respectively.
The friction coefficients has been already derived (equation
\ref{gx}):
\begin{equation}
\gamma_z=2\gamma_{x,y}
\end{equation}
Indeed cooling along an axe in the $x-y$ plane results in the action
of two counter-propagating beams four times less coupled than the
single upward laser beam. The same argument holds for the absorption
term of the diffusion coefficient:
\begin{equation}
D^{abs}_z=2D^{abs}_{x,y}
\end{equation}
The spontaneous emission contribution in the diffusion coefficient
can be derived from the differential cross-section
$\textrm{d}\sigma/\textrm{d}\Omega$ of the emitting dipole
\cite{jackson}. With a strong biased magnetic field along the
vertical direction, this calculation is particularly simple as
$e^+_z$ is the only quasi resonant state. Hence
\begin{equation}
\textrm{d}\sigma/\textrm{d}\Omega\propto(1+\cos{\phi}^2)
\end{equation} $\phi$ is the angle between the vertical axe
and the direction of observation. After a straightforward
integration, one finds a contribution again two times larger along
the vertical axe:
\begin{equation}
D^{spo}_z=2D^{spo}_{x,y}
\end{equation}
From those considerations, the temperature is expected to be
isotropic as observed experimentally (see figures \ref{TempNLC}a and
\ref{TempNLC}b).

In the so-called regime (I), the minimum temperature is given by the
semiclassical Doppler theory:
\begin{equation}
T=N_R\frac{\hbar \Gamma}{2k_B}\sqrt{s} \label{TempF}\end{equation}
Where $N_R$ is a numerical factor which should be close to two
\cite{LoftusPRA}. This solution is represented in figure
\ref{TempNLC} by a dashed line nicely matching the experimental data
for $s>8$ but with $N_R=1.2$. Similar results, \emph{i.e.} with
unexpected low $N_R$ values, have been found in \cite{LoftusPRA}.
For $s\leq 8$ we observed a plateau in the final temperature
slightly higher than the low saturation theoretical prediction
\cite{Castin}. We cannot explain why the temperature does not
decrease further down as reported in \cite{LoftusPRA}. For
quantitative comparison with the theory, more detailed studies in a
horizontal 1D molasses are required.

\subsection{Conclusions}
\label{cclMC}

Cooling of Strontium atoms using the intercombination line is an
efficient technique to reach the recoil temperature in three
dimensions by optical methods. Unfortunately loading from a thermal
beam cannot be done directly with a single frequency laser because
of the narrow velocity capture range. We have shown experimentally
that more than $50\%$ of the atoms initially in a blue MOT on the
dipole-allowed transition are recaptured in the red MOT using a
frequency-broadened spectrum. Using a simple model, we conclude that
the transfer is limited by the size of the laser beam. If the total
power of the beams at $689\U{nm}$ was higher, transfer rates up to
$90\%$ could be expected by tripling our laser beam size. The final
temperature in the broadband regime is found to be as low as
$2.5\U{\mu K}$, \emph{i.e.} only 5 times larger than the photon
recoil temperature. The gain in temperature by comparison to the
blue MOT ($1-10\U{mK}$) is appreciable. So in absence of strong
requirements on the temperature, broadband cooling is very efficient
and reasonably fast (less than $100\U{ms}$). The requirements for
the frequency noise of the laser are also much less stringent than
for single frequency cooling.

Using a subsequent single frequency cooling stage, it is possible to
reduce the temperature down to $600\U{nK}$, slightly above the
photon recoil temperature. Analyzing the large detuning regime, we
particularly focus our studies on the comparison between vertical
and horizontal directions. We show how gravity indirectly influences
the horizontal parameters of the steady state MOT and find that the
trapping potential remains harmonic along all directions, but with
an anisotropy.

Gravity has a major impact on the MOT as it counterbalances the
laser pressure of the upward laser (making the steady state
independent of the detuning). We show that gravity thus affects the
final temperature, which remains isotropic, despite different
cooling dynamics along the vertical and horizontal directions.

\section{Acknowledgments}

The authors wish to thank J.-C. Bernard and J.-C. Bery for valuable
technical assistances. This research is financially supported by the
CNRS (Centre National de la Recherche Scientifique) and the former
BNM (Bureau National de M\'etrologie) actually LNE (Laboratoire
national de m\'etrologie et d'essais) contract N$^{\circ}$ 03 3 005.

\newpage

\end{document}